\documentclass[11pt]{article}
\usepackage{amsfonts}
\usepackage{mathrsfs}
\usepackage{amsfonts,amssymb,mathrsfs}
\usepackage{amsmath,amscd}
\usepackage[dvips]{graphicx}
\usepackage{color}
\usepackage[all]{xy}
\usepackage{graphicx}
\usepackage{mathptm,pslatex}

\definecolor{red}{rgb}{1,0,0}
\definecolor{blue}{rgb}{0,0,1}
\definecolor{darkblue}{rgb}{0,0,0.4}

\begin{document}
%\hspace{11cm}{Primary version} \sloppy
%\newcommand{\dickebox}{{\vrule height5pt width5pt depth0pt}}
%\pagestyle{fancy} \fancyhf{} \fancyfoot[CE,CO]{{\thepage}}
%\fancyhead[LE]{\nouppercase{\rightmark}}
%\fancyhead[RO]{\nouppercase{\leftmark}}
\newtheorem{Def}{Definition}[section]
\newtheorem{exe}{Example}[section]
\newtheorem{prop}[Def]{Proposition}
\newtheorem{theo}[Def]{Theorem}
\newtheorem{lem}[Def]{Lemma}
\newtheorem{rem}{\noindent\mbox{Remark}}[section]
\newtheorem{coro}[Def]{Corollary}
 \newcommand{\Ker}{\rm Ker}
  \newcommand{\Soc}{\rm Soc}
 \newcommand{\lra}{\longrightarrow}
 \newcommand{\ra}{\rightarrow}
 \newcommand{\add}{{\rm add\, }}
\newcommand{\gd}{{\rm gl.dim\, }}
\newcommand{\End}{{\rm End\, }}
\newcommand{\overpr}{$\hfill\square$}
\newcommand{\rad}{{\rm rad\,}}
\newcommand{\soc}{{\rm soc\,}}
\renewcommand{\top}{{\rm top\,}}
\newcommand{\fdim}{{\rm fin.dim}\,}
\newcommand{\fidim}{{\rm fin.inj.dim}\,}
\newcommand{\gldim}{{\rm gl.dim}\,}
\newcommand{\cpx}[1]{#1^{\bullet}}
\newcommand{\D}[1]{{\mathscr D}(#1)}
\newcommand{\Dz}[1]{{\rm D}^+(#1)}
\newcommand{\Df}[1]{{\rm D}^-(#1)}
\newcommand{\Db}[1]{{\mathscr D}^b(#1)}
\newcommand{\C}[1]{{\mathscr C}(#1)}
\newcommand{\Cz}[1]{{\rm C}^+(#1)}
\newcommand{\Cf}[1]{{\rm C}^-(#1)}
\newcommand{\Cb}[1]{{\mathscr C}^b(#1)}
\newcommand{\K}[1]{{\mathscr K}(#1)}
\newcommand{\Kz}[1]{{\rm K}^+(#1)}
\newcommand{\Kf}[1]{{\rm K}^-(#1)}
\newcommand{\Kb}[1]{{\mathscr K}^b(#1)}
\newcommand{\modcat}[1]{#1\mbox{{\rm -mod}}}
\newcommand{\stmodcat}[1]{#1\mbox{{\rm -{\underline{mod}}}}}
\newcommand{\pmodcat}[1]{#1\mbox{{\rm -proj}}}
\newcommand{\imodcat}[1]{#1\mbox{{\rm -inj}}}
\newcommand{\opp}{^{\rm op}}
\newcommand{\otimesL}{\otimes^{\rm\bf L}}
\newcommand{\rHom}{{\rm\bf R}{\rm Hom}\,}
\newcommand{\pd}{{\rm pd}}
\newcommand{\Hom}{{\rm Hom \, }}
\newcommand{\Coker}{{\rm coker}\,\,}
\newcommand{\Ext}{{\rm Ext}}
\newcommand{\im}{{\rm im}}
\newcommand{\Cone}{{\rm Cone}}
\newcommand{\fini}{{finitistic}}
\newcommand{\proj}{{\rm  proj}}
\newcommand{\al}{{\alpha}}
\newcommand{\tr}{{ tr}}
\newcommand{\Tor}{{\rm  Tor}}
\newcommand{\ad}{{\rm  add}}
\newcommand{\StHom}{{\rm \underline{Hom} \, }}
\pagenumbering{arabic}{\Large \bf
\begin{center}
%%%%%%%%%%%%%%%%%%%%%%%%%%%%title%%%%%%%%%%%%%%%%%%%%%%%%%%%%
Construction of mutually unbiased maximally entangled bases in $\mathbb{C}^{2^s}\otimes\mathbb{C}^{2^s}$ by using Galois rings
%%%%%%%%%%%%%%%%%%%%%%%%%%%%title%%%%%%%%%%%%%%%%%%%%%%%%%%%
\end{center}}
\centerline{ {\sc Dengming xu$^*$ }}

\abstract{Mutually unbiased bases plays a central role in quantum mechanics and quantum information processing. As an important class  of  mutually unbiased bases,
mutually unbiased maximally entangled bases (MUMEBs) in bipartite systems
  have attracted much attention in recent years. In the paper, we try to  construct MUMEBs in $\mathbb{C}^{2^s}\otimes\mathbb{C}^{2^s}$ by using  Galois rings, which is different from the work in \cite{xu2}, where  finite fields  are used.
As applications,  we obtain several new types of MUMEBs  in $\mathbb{C}^{2^s}\otimes\mathbb{C}^{2^s}$  and prove that  $M(2^s,2^s)\geq 3(2^s-1)$, which raises the lower  bound of $M(2^s,2^s)$ given in \cite{xu}. }

\vskip -1.4cm
\renewcommand{\thefootnote}{\alph{footnote}}
\renewcommand{\thefootnote}{\alph{footnote}}
\setcounter{footnote}{-1} \footnote{$^*$Sino-European Institute of Aviation Engineering,
 Civil Aviation University of China, 300300 Tianjin,
People's Republic of  China.
}

\renewcommand{\thefootnote}{\alph{footnote}}
\setcounter{footnote}{-1} 
\renewcommand{\thefootnote}{\alph{footnote}}
\setcounter{footnote}{-1} \footnote{ E-mail:
xudeng17@163.com,\ Tel: 0086-22-24092750.}

\section{Introduction}

Mutually unbiased bases (MUBs) and maximally entangled states play a central role in quantum mechanics   and quantum information processing, such as  qutntum key distribution, cryptographic protocols, mean king's problem and quantum teleportation and super dense coding, see \cite{zyc, feng} for detalis.  As an important class of  mutually unbiased bases,
mutually unbiased maximally entangled bases (MUMEBs) in bipartite systems
   also    have a close relation with  unitary $2$-design (\cite{sco}) and mutually unbiased unitary bases (\cite{sha}). Moreover, MUMEBs  can be used to construct MUBs  in a Hilbert space of composite order (\cite{fei1}).
    In recent years, construction of MUMEBs   has attracted much attention (\cite{cc,feng,xu,xu2}).

Let $d\in \mathbb{N}$ such that $d\geq 2$.  Two orthonormal bases $\mathcal{B}_1=\{|\phi_i\rangle\mid 1\leq i\leq d\}$ and
 $\mathcal{B}_2=\{|\psi_i\rangle\mid 1\leq i\leq d\}$ of $\mathbb{C}^d$ are said to be  mutually unbiased if
 $$\forall 1\leq i,j\leq d,\quad |\langle\phi_i|\psi_j\rangle|=\dfrac{1}{\sqrt{d}}.$$

 \noindent A set of orthonormal bases $\mathcal{B}_1,\mathcal{B}_2,\cdots,\mathcal{B}_m$
in $\mathbb{C}^d$ is called a set of mutually unbiased bases (MUBs) if every pair in the set
  is mutually unbiased.  Let $N(d)$ denote the maximum cardinality of any set of MUBs in  $\mathbb{C}^d$. It was proved  that  $N(d) = d + 1$ if
$d$ is a prime power (\cite{Vat,Rot,Fie}). However, little has been known when $d$ is a composite number, the exact value of  $ N(6)$ is still unknown.

  Suppose $d\geq 2$. A (pure) maximally entangled state in $\mathbb{C}^d\otimes\mathbb{C}^{d}$
 can be written as
 $$|\psi\rangle=\dfrac{1}{\sqrt{d}}\displaystyle\sum_{i=1}^{d}|e_i\rangle\otimes U|e_i\rangle,$$
 where $\{|e_i\rangle\mid 1\leq i\leq d\}$ is an  orthonormal  bases of  $\mathbb{C}^d$ and $U$ is a  unitary operator on $ \mathbb{C}^d$.
  A base $\mathscr{B}$ of $\mathbb{C}^d\otimes\mathbb{C}^{d}$ is called a maximally entangled base (MEB) if each element in  $\mathscr{B}$  is a   maximally entangled state.  Let $\mathscr{A}=\{\mathcal{B}_1,\mathcal{B}_2,\cdots,\mathcal{B}_m\}$  be a set  of orthonormal MEBs in  $\mathbb{C}^d\otimes\mathbb{C}^{d}$.
 We call  $\mathscr{A}$ a set of mutually unbiased maximally entangled bases (MUMEBs) if   every pair in $\{\mathcal{B}_1,\mathcal{B}_2,\cdots,\mathcal{B}_m\}$ is mutually unbiased.
 Let $M(d,d)$ be the maximal cardinality of any set of mutually unbiased maximally entangled bases
 in   $\mathbb{C}^d\otimes\mathbb{C}^{d}$.   It is obvious that $M(d,d)\leq N(d)$.
 In \cite{sco}, the author proved  that $M(d,d)\leq d^2-1$ when $d$ is a prime, and $ M(d,d)=d^2-1$ for  $d=2,3,5,7,11$.
  However, it is still unclear if this is true for all prime numbers.  In \cite{feng, fei1}, the authors provided  general methods to construct   MUMEBs.
In \cite{xu}, it was proved that $M(d,d)\geq 2(d-1)$  for any prime power number $d$.
In addition, the authors in \cite{cc} constructed   MUMEBs in $\mathbb{C}^d\otimes\mathbb{C}^d$ when $d$ is a composite number.

Let $q$ be  an odd prime power, the author in \cite{xu2} proved that  $M(q,q)\geq \frac{q^2-1}{2}$ by constructing  some special subsets in the special linear group
$SL(2,\mathbb{F}_q)$. Unfortunately, the construction is not applicable when $q=2^s$. Let $s\geq 2$.  Inspired by the work in \cite{xu2},
by using  Galois rings instead of finite fields used in \cite{xu2}, we construct MUMEBs in  $\mathbb{C}^{2^s}\otimes\mathbb{C}^{2^s}$
  through trace-zero excluded subsets of the special linear group $SL(2, \mathbb{F}_{2^s})$.
 Namely,  a non-empty subset $\mathcal{C}$ of $ SL(2,\mathbb{F}_{2^s})$ is called
a {\bf trace-zero excluded subset} if for  any different $A$ and $B$ in $\mathcal{C}$, the trace  of $A^{-1}B$ is nonzero.
Based on the definition, by means of basic results on  the Galois ring  $GR(4,4^s)$, we obtain the following result (see  Theorem \ref{theo2}).

\noindent{\it  {\bf Theorem A}
Suppose that $\mathcal{C}$ is a trace-zero excluded subset of $SL(2,\mathbb{F}_{2^s})$.
Then $(\Phi_{A})_{A\in \mathcal{C}}$ is a  set of MUMEBs in $\mathbb{C}^{2^s}\otimes\mathbb{C}^{2^s}$.}

Theorem A provides a new  method to construct MUMEBs in $\mathbb{C}^{2^s}\otimes\mathbb{C}^{2^s}$. It gives us   a possibility to higher the lower bound of $M(2^s,2^s)$.
 Based on  Theorem A, we construct new  types of   MUMEBs in $\mathbb{C}^{2^s}\otimes\mathbb{C}^{2^s}$ by seeking trace-zero excluded subsets in
$SL(2,\mathbb{F}_{2^s})$.
As an application of the theorem, we prove that $M(2^s,2^s)\geq 3(2^s-1)$ (see Proposition \ref{propex2}), which raises the lower  bound of $M(2^s,2^s)$ given in \cite{xu}.

The paper is organized  as follows. In Section 2,  we first introduce  basic  definitions and results about Galois rings, and then briefly  recall  a  general  construction of  MUMEBs in $\mathbb{C}^{2^s}\otimes\mathbb{C}^{2^s}$ given in \cite{xu}, finally we  show  how to construct unitary matrices  from $SL(2,\mathbb{F}_{2^s})$. In Section 3, we   give a proof of Theorem A and     construct   MUMEBs  in $\mathbb{C}^{2^s}\otimes\mathbb{C}^{2^s}$   by seeking trace-zero excluded subsets of $SL(2,\mathbb{F}_{2^s})$. In Section 4, we give  conclusions of the paper.

\section{Preliminaries}
\subsection{An introduction to basic facts about Galois rings}

First, we  recall basic definitions and facts about  Galois rings from  \cite{Car,Wan}.

Let $h_2(x)\in\mathbb{Z}_2[X]$ be a primitive  polynomial of degree $s\geq 2$. Then there is a  unique monic polynomial $h(x)\in\mathbb{Z}_4[X]$
of degree $s$ such that $h(x)\equiv h_2(x)$ (mod $2$) and $h(x)$ divides $x^{2^s-1}-1$ (mod $ 4$). Let $\xi$ be a root of $h(x)$ such that $\xi^{2^s-1}=1$.
 Then the {\bf Galois ring} $GR(4,4^s)$ is defined by $\mathscr{R}=\mathbb{Z}_4(\xi).$ Let $\mathcal{T}_s=\{0, 1,\xi,\xi^2,\cdots, \xi^{2^s-2}\}$. Then
$$ \mathbb{Z}_4(\xi)=\{a+2b\mid (a,b)\in\mathcal{T}_s\times\mathcal{T}_s\}.$$
The Galois ring $\mathscr{R}$ has a unique  maximal ideal $2\mathscr{R}$ and  the residue field $\mathscr{R}/2\mathscr{R}$ is isomorphic to $\mathbb{F}_{2^s},$ we will identify $\mathscr{R}/2\mathscr{R}$ with $\mathbb{F}_{2^s}.$
The {\it Frobenius map} $f$ from $\mathscr{R}$ to $\mathscr{R}$ is defined by
 $$\forall \   a,b\in\mathcal{T}_s, \ f(a+2b)=a^2+2b^2.$$
 The {\it relative trace } $\tr$ from $\mathscr{R}$ to $\mathbb{Z}_4$ is defined by
$$\forall \   c\in \mathscr{R}, \quad \tr(c)=c+f(c)+f^2(c)+\cdots+f^{s-1}(c).$$
The
 {\it additive character} $\lambda$ of $(\mathscr{R},+)$ is defined by£º $\forall \   x\in \mathscr{R}, \lambda(x)=i^{\tr(x)}.$
It is easy to see from the definition that
$$\forall \   a,b\in\mathcal{T}_s, \quad \lambda(a)=\lambda(a^2), \quad \lambda(2b)=\lambda(2b^2),$$
we will use these facts frequently in  the  rest of the paper.

Let $a,b\in \mathcal{T}_s$. Since $(a^{2^{s-1}})^2=a$, we set $\sqrt{a}=a^{2^{s-1}}$. Then  one can check that $a+b+2\sqrt{ab}\in\mathcal{T}_s$ and we  write $$a+b=\underset{\in\mathcal{T}_s}{\underbrace{a+b+2\sqrt{ab}}}+2(\underset{\in\mathcal{T}_s}{\underbrace{\sqrt{ab}}}).$$
Define $a\oplus b=a+b+2\sqrt{ab}.$ Then we have $$(a\oplus b)^2=(a+b)^2, \qquad 2(a\oplus b)=2(a+b).$$
In this way, $(\mathcal{T}_s, \oplus, \cdot)$ forms a field. Let $\mu$ be the canonical map from $\mathscr{R}\rightarrow\mathscr{R}/2\mathscr{R}\simeq \mathbb{F}_{2^s}$, we  get an isomorphism of fields:
 $$\begin{array}{cccc}
    \phi: & (\mathcal{T}_s, \oplus, \cdot) & \rightarrow & (\mathbb{F}_{2^s}, +, \cdot) \\
   & x& \mapsto &\mu(x)
   \end{array}
$$
Based on the isomorphism, for any $x\in \mathbb{F}_{2^s}$,  $\lambda(\phi^{-1}(x))$ will be denoted  by $\lambda(x)$ for abbreviation.  In this way,
for any $a,b\in \mathbb{F}_{2^s},$  $\lambda(\phi^{-1}(a+b)$ will be denoted by  $\lambda(a\oplus b)$.

For the proof of the main results, we need the following lemma.

\begin{lem}\label{lem21} \cite[Lemma 3]{Car} Keep these notations as above. Then the following is true.
\begin{itemize}
  \item [(1)]Set $\Gamma(r)=\displaystyle\sum_{x\in\mathcal{T}_s}\lambda(rx)$ for each $r\in\mathscr{R}$.
  Let $a,b\in \mathcal{T}_s$ with $a\neq 0$. Then $\Gamma(a+2b)=\Gamma(1)\lambda(-a^{-1}b)$ and $|\Gamma(1)|=\sqrt{2^s}$.
  \item [(2)] Let $r\in\mathscr{R}$. Then $$\big|\displaystyle\sum_{x\in\mathcal{T}_s}\lambda(rx)\big|=
\left\{\begin{array}{cl}
  0, &  r\in 2\mathcal{T}_s,r\not=0; \\
  2^s, & r=0; \\
  \sqrt{2^s}, & else.
\end{array}\right.
$$
\end{itemize}

\end{lem}

\subsection{General construction of  mutually unbiased maximally entangled bases by using Galois rings.}

From now on, let $s\in\mathbb{N}$ such that  $s\geq 2$. Set $q=2^s$ and  $F=\mathbb{F}_{2^s}$. Let $\xi$ be a  fixed primitive element in $F$, we order the elements in $F$  as
$F=\{0,1,\xi,\xi^2,\cdots,\xi^{2^s-2}\}.$
  Fix  an  orthonormal basis $\{|e_r\rangle: r\in F\}$ of $\mathbb{C}^q$.
Define Pauli operators
$$H_{\xi,\eta}=\displaystyle\sum_{r\in F}\lambda(2(r\xi))|e_{r+\eta}\rangle\langle e_r|,\qquad {\xi,\eta\in F}$$
Given a  unitary matrix $U$ in $M_{q}(\mathbb{C})$, we know that
$\{|Ue_{r}\rangle: r\in F\}$ is again an orthonormal basis of $\mathbb{C}^{q}$.
Applying $H_{\xi,\eta}\otimes I_{q}$ on the following  maximally entangled state $|\psi_U\rangle$,
 $$|\psi_U\rangle=\dfrac{1}{\sqrt{q}}\displaystyle\sum_{r\in F}|e_r\rangle\otimes U|e_{r}\rangle. $$
 we get  $q^2$ maximally entangled states:
$$(H_{\xi,\eta}\otimes I_{q})|\psi_U\rangle=\dfrac{1}{\sqrt{q}}\displaystyle\sum_{r\in F}\lambda (2r\xi)|e_{r+\eta}\rangle \otimes U|e_{r}\rangle
\quad \xi,\eta\in F.$$
Set $\Psi_U=\{(H_{\xi,\eta}\otimes I_{q})|\psi_U\rangle\mid \xi,\eta\in F\}.$

Now we can state the following lemma.

\begin{lem}\cite[Lemma 4.2]{xu}
\label{lem22}
(1)  For any unitary matrix $U$ in  ${M}_{q}(\mathbb{C})$, $\Psi_U$ is an orthonormal MEB
 in $\mathbb{C}^{q}\otimes\mathbb{C}^{q}$.

\noindent  (2) Given two unitary matrices $U,V$  in  ${M}_{q}(\mathbb{C})$, set $W=U^*V=(w_{i,j})$. Then $\Psi_U$ and $\Psi_V$ in
 $\mathbb{C}^{q}\otimes\mathbb{C}^{q}$ are mutually unbiased if and only if
\begin{equation}
\forall \   \xi,\eta \in F,\quad |\displaystyle\sum_{r\in F}\lambda(2r\xi)w_{r,r+\eta}|=1.
 \tag{4.1}
\end{equation}
 \end{lem}

\subsection{Unitary matrices constructed from $SL(2,F)$}

   The work of this subsection is inspired by \cite{app, xu2}. We will use the Galois ring $\mathscr{R}$ instead of $\mathbb{Z}_q$ used in \cite{app} and the finite field $\mathbb{F}_q$ used in \cite{xu2}. Recall that  $$SL(2,F)=\bigg\{\left[\begin{smallmatrix}
  \alpha & \beta\\
  \gamma&\delta
\end{smallmatrix}\right]\in M_2(F)\mid \alpha\delta-\beta\gamma=1\bigg\}.$$

Let
$A=
\left[\begin{smallmatrix}
  \alpha & \beta\\
  \gamma&\delta
\end{smallmatrix}\right]
\in SL(2,F)$.

\begin{itemize}
  \item  Suppose $\beta\neq0$. Define  $V_A$ as:
$$\forall \   m,n\in F, \qquad  (V_A)_{m,n}=\frac{1}{\sqrt{q}}\lambda({\beta^{-1}}(\alpha n^2+2mn+\delta m^2)).$$
  \item Suppose $\beta=0$. Then $\delta\neq 0$. Let
  $$L=\left[\begin{matrix}
  0 & 1\\
  1&0
\end{matrix}\right], \quad
K=\left[\begin{matrix}
  \gamma&\delta\\
  \alpha & 0
\end{matrix}\right].$$  Then $L,K\in  SL(2,F)$ and  $A=LK$.  We define $V_A=V_LV_K$.
\end{itemize}

\begin{lem}\label{lem23}  Let
$A=
\left[\begin{smallmatrix}
  \alpha & \beta\\
  \gamma&\delta
\end{smallmatrix}\right]
\in SL(2,F)$.
\begin{itemize}
  \item [$(1)$] If $\beta=0$, then $$\forall \   m,n\in F, \quad   (V_A)_{m,n}=\left\{
\begin{array}{ll}
 \lambda({mn\gamma})& m=\alpha n\\
  0& else
\end{array}\right.$$ Moreover, if $\gamma=0$, then $V_A$ is the permutation matrix defined in \cite{feng}.
In particular, $V_{I_2}=I_q$.
  \item [$(2)$] $V_A$ is a unitary matrix.
\end{itemize}

\end{lem}

\noindent {\bf Proof} (1) If $\beta=0, $ then $\alpha\neq 0$ and $\delta=\alpha^{-1}$.
Let $m,n\in F$. Then $$\begin{array}{lll}(V_A)_{m,n}=\displaystyle\sum_{k\in F} (V_L)_{m,k}(V_K)_{k,n}
&=&\dfrac{1}{q}\displaystyle\sum_{k\in F}  \lambda(2mk+\delta^{-1}(\gamma n^2+2kn))\\
&=&\dfrac{1}{q}\lambda(\alpha\gamma n^2)\displaystyle\sum_{k\in F}  \lambda(2k(m+\alpha n))\quad (\delta=\alpha^{-1})\\
&=&\dfrac{1}{q}\lambda(\alpha\gamma n^2)\displaystyle\sum_{k\in F}  \lambda(2k(m\oplus\alpha n))
\end{array}
$$
 Then the result follows from Lemma \ref{lem21}

(2) If $\beta=0,$ then we get from (1) that $V_A$ is unitary. Now suppose that $\beta\neq 0.$ Let $m,n\in F$. Then
 we have
 $$\begin{array}{lll}(V_A^*V_A)_{m,n}=\displaystyle\sum_{k\in F} (\overline{V_A})_{k,m}(V_K)_{k,n}
& =&\dfrac{1}{q}\displaystyle\sum_{k\in F}\lambda ( {\beta^{-1}}(\alpha n^2+2nk+\delta k^2)-\beta^{-1}(\alpha m^2+2km+\delta k^2))\\
 &=&\dfrac{1}{q}\lambda (\beta^{-1}\alpha(n^2-m^2)\displaystyle\sum_{k\in F}\lambda(2k\beta^{-1}(n-m))\\
 &=&\dfrac{1}{q}\lambda (\beta^{-1}\alpha(n^2-m^2)\displaystyle\sum_{k\in F}\lambda(2k\beta^{-1}(n\oplus m))
 \end{array}$$
It follows from Lemma \ref{lem21} that  $V_A$ is unitary. \qquad $\square $

The rest of this section will  not be used in our proofs,
we  show it here and hope that it may be of its own interest.
Let $a\in F.$  Define two matrices $X_a$ and $Z_a$ in ${M}_{q}(\mathbb{C})$ as follows:
$$\forall \   m,n\in F,  \qquad
(X_a)_{m,n}=
\left\{
\begin{array}{ll}
 1 & if \ m=n+a\\
  0& else
\end{array}\right.\qquad \mbox{and } \qquad
(Z_a)_{m,n}=\left\{
\begin{array}{ll}
  \lambda (2ma)& if\  m=n\\
  0& else
\end{array}\right.$$
Then $X_a^*=X_a$ and $Z_a^*=Z_{a}$. Let $v=(a,b)\in F^2$. One can check $Z_bX_a=\lambda(2ab)X_aZ_b$. Define $D_{v}=\lambda({ab})X_aZ_b$.
Then $D_v^*=D_{v}$ and
$$\forall \   m,n\in F,  \qquad
(D_{v})_{m,n}=\left\{
\begin{array}{ll}
 \lambda({ab}+2bn)& m=n+a\\
  0& else
\end{array}\right.
$$

\begin{lem}\label{lem24}
 Keep the notations as above.   Then $\forall \   A\in SL(2,F), \forall \   v\in F^2, V_AD_vV_A^*=\pm D_{Av}$.
\end{lem}

\noindent {\bf Proof \ } Let $A=
\left[\begin{smallmatrix}
  \alpha & \beta\\
  \gamma&\delta
\end{smallmatrix}\right]
\in SL(2,F)$.

 (1) \ Suppose $\beta\neq 0$. Set $v=(a,b).$ By definition, we have  $$\forall \   m,n\in F,  \quad(V_A)_{m,n}=\frac{1}{\sqrt{q}}\lambda({\beta^{-1}}(\delta m^2+2mn+\alpha n^2)),\quad
(D_{v})_{m,n}=\left\{
\begin{array}{ll}
 \lambda(ab+2bn)& m=n+a\\
  0& else
\end{array}\right.
$$
Then for each $(m,n)\in F^2$, we have
$$(V_A D_{v})_{m,n}=\displaystyle\sum_{k\in F} (V_A)_{m,k}(D_{v})_{k,n}=(V_A)_{m,n+ a}(D_{v})_{n+ a,n}=\frac{1}{\sqrt{q}}\lambda(ab+2bn+{\beta^{-1}}(\delta m^2+2(n\oplus a)m+\alpha (n\oplus a)^2))$$
$$=\frac{1}{\sqrt{q}}\lambda(ab+2bn+{\beta^{-1}}(\delta m^2+\alpha n^2+\alpha a^2+2nm+2am+2\alpha na)$$
Similarly, $$(D_{v}V_A)_{m,n}=\displaystyle\sum_{k\in F} (D_{v})_{m,k}(V_A)_{k,n}=(D_{v})_{m,m+a}(V_A)_{m+ a,n}=\frac{1}{\sqrt{q}}\lambda(-ab+2bm+{\beta^{-1}}(\delta(m\oplus a)^2+2(m\oplus a)n+\alpha n^2)).$$
$$=\frac{1}{\sqrt{q}}\lambda(-ab+2bm+{\beta^{-1}}(\delta m^2+\delta a^2+\alpha n^2+2\delta ma+2mn+2an))$$
Set $Av=(\alpha a+\beta b,\gamma a+\delta b)=(a',b').$ Recall that  in $\mathscr{R}$, we have equations: $2a'=2(\alpha a\oplus\beta b), 2b'= (\gamma a\oplus\delta b), 2(a')^2=(\gamma a\oplus\delta b)^2$  and $(\alpha a\oplus\beta b)(\gamma a\oplus\delta b))=(\alpha a+\beta b+2\sqrt{\alpha\beta ab})(\gamma a+\delta b+2\sqrt{\gamma\delta ab}))$.
Then we have
$$\begin{array}{lll}(D_{Av})_{m,k}(V_A)_{k,n}
&=&\dfrac{1}{\sqrt{q}}\lambda(-(\alpha a\oplus\beta b)(\gamma a\oplus \delta b)+2b'm+{\beta^{-1}}(\delta m^2+\delta a'^2+\alpha n^2+2\delta ma'+2mn+2a'n))\\
&=& \dfrac{1}{\sqrt{q}}\lambda(2\sqrt{\alpha\beta ab}(\gamma a+\delta b)+2\sqrt{\gamma\delta ab}(\alpha a+\beta b))\times\\
&& \lambda(-(\alpha a+\beta b)(\gamma a+\delta b)+2b'm+{\beta^{-1}}(\delta m^2+\delta a'^2+\alpha n^2+2\delta ma'+2mn+2a'n)).
\end{array}$$
Similar to the proof of Lemma 3.2 in \cite{xu2}, one can check that  $$(D_{Av}V_A)_{m,n}=(V_A D_v)_{m,n}\ \lambda(2\sqrt{\alpha\beta ab}(\gamma a+\delta b)+2\sqrt{\gamma\delta ab}(\alpha a+\beta b))$$ by a careful calculation.  As a result, $$V_A D_v V_A^*=\lambda(2\sqrt{\alpha\beta ab}(\gamma a+\delta b)+2\sqrt{\gamma\delta ab}(\alpha a+\beta b))\  D_{Av},$$ where $\lambda(2\sqrt{\alpha\beta ab}(\gamma a+\delta b)+2\sqrt{\gamma\delta ab}(\alpha a+\beta b)) =\pm1.$

(2) \ Suppose that $\beta=0$. Then it follows from (1) and the definition of $V_A$ that the result is also true. \quad
$\square$

\section{Proof of the  Main results}
\subsection{The  sufficient condition}
For simplicity, we denote $\Phi_A=\Phi_{V_A}$ for each $A\in SL(2,F)$. Now we study when $\Phi_A$ and $\Phi_B$ are mutually unbiased for
$A,B\in SL(2,F).$

\begin{lem}  \label{propth1}Let
$A=
\left[\begin{smallmatrix}
  \alpha_1 & 0\\
  \gamma_1&\alpha_1^{-1}
\end{smallmatrix}\right]
$ and $
B=
\left[\begin{smallmatrix}
  \alpha_2 &0\\
  \gamma_2&\alpha_2^{-1}
\end{smallmatrix}\right]
$ in $SL(2,F)$. If  $trace(A^{-1}B)\neq0$, then $\Phi_A$ and $\Phi_B$ are mutually unbiased.
\end{lem}

\noindent {\bf Proof} Let $m,n\in F.$ Then we have by Lemma \ref{lem23} that
 $$(V_A^*V_B)_{m,n}=\displaystyle\sum_{k\in F}\overline{(V_A)_{k,m}}(V_B)_{k,n}
=\left\{\begin{array}{cc}
\overline{(V_A)_{\alpha_1m,m}}(V_B)_{\alpha_2n,n}=\lambda(\alpha_2n^2\gamma_2-\alpha_1\gamma_1m^2),& m=\alpha_1^{-1}\alpha_2n\\
0, &\mbox{else}
\end{array}\right. $$
If $trace(A^{-1}B)\neq0$, then $\alpha_1^{-1}\alpha_2\neq 1$. It follows $1+\alpha_1^{-1}\alpha_2\neq 0.$ Let $\xi,\eta\in F$. Then
$$\bigg|\displaystyle\sum_{r\in F}\lambda(2r\xi)(V_A^*V_B)_{r,r+\eta}\bigg|=\big|\lambda(2r\xi)\ \ \lambda(\alpha_2(r\oplus \eta)^2\gamma_2-\alpha_1\gamma_1r^2)\big|=1. \qquad \big(r=(1+\alpha_1^{-1}\alpha_2)^{-1}\alpha_1^{-1}\alpha_2\eta\big)$$
It follows from Lemma \ref{lem22}
 that $\Phi_A$ and $\Phi_B$ are mutually unbiased. \qquad $\square$

\begin{lem}   \label{propth2}Let
$A=
\left[\begin{smallmatrix}
  \alpha_1 & 0\\
  \gamma_1&\alpha_1^{-1}
\end{smallmatrix}\right]
$ and $
B=
\left[\begin{smallmatrix}
  \alpha_2 & \beta_2\\
  \gamma_2&\delta_2
\end{smallmatrix}\right]
$ in $SL(2,F)$ such that $\beta_2\neq 0$. If $trace(A^{-1}B)\neq0$, then $\Phi_A$ and $\Phi_B$ are mutually unbiased.
\end{lem}

\noindent {\bf Proof} Let $m,n\in F.$ It follows from Lemma \ref{lem23} that
$$(V_A^*V_B)_{m,n}=\displaystyle\sum_{k\in F}\overline{(V_A)_{k,m}}(V_B)_{k,n}=\overline{(V_A)_{\alpha_1m,m}}(V_B)_{\alpha_1m,n}
=\frac{1}{\sqrt{q}}\lambda(\beta_2^{-1}(\alpha_2 n^2+2\alpha_1mn+\delta_2(\alpha_1m)^2)-\alpha_1m^2\gamma_1)$$
Let $\xi,\eta\in F$. Then
$$\begin{array}{lll}\displaystyle\sum_{r\in F}\lambda(2r\xi)(V_A^*V_B)_{r,r+\eta}&=&
\dfrac{1}{\sqrt{q}}\displaystyle\sum_{r\in F}\lambda(2r\xi+\beta_2^{-1}(\alpha_2 (r\oplus \eta)^2+2\alpha_1r(r\oplus \eta)+\delta_2(\alpha_1r)^2)-\alpha_1r^2\gamma_1)\\&=&\dfrac{\lambda(\beta_2^{-1}\alpha_2\eta^2)}{\sqrt{q}}\displaystyle\sum_{r\in F}
\lambda(r^2(\beta_2^{-1}\alpha_2+ \beta_2^{-1}\delta_2\alpha_1^2-\alpha_1\gamma_1)+2r^2(\xi^2+\beta_2^{-2}\alpha_2^2\eta^2+\beta_2^{-1}\alpha_1+\beta_2^{-2}\alpha_1^2\eta^2))
\\&=&\dfrac{\lambda(\beta_2^{-1}\alpha_2\eta^2)}{\sqrt{q}}\displaystyle\sum_{r\in F}
\lambda(r(\beta_2^{-1}\alpha_2+ \beta_2^{-1}\delta_2\alpha_1^2-\alpha_1\gamma_1)+2(\xi^2+\beta_2^{-2}\alpha_2^2\eta^2+\beta_2^{-1}\alpha_1+\beta_2^{-2}\alpha_1^2\eta^2)))\\
&=&\dfrac{\lambda(\beta_2^{-1}\alpha_2\eta^2)}{\sqrt{q}}\displaystyle\sum_{r\in F}
\lambda(r((\beta_2^{-1}\alpha_2\oplus \beta_2^{-1}\delta_2\alpha_1^2\oplus\alpha_1\gamma_1)+2c))
\end{array}$$
for some $c\in \mathcal{T}_s$. It follows from Lemmas \ref{lem21} and \ref{lem22}
 that  $\Phi_A$ and $\Phi_B$ are mutually unbiased if and only if  $\big|\displaystyle\sum_{r\in F}\lambda(2r\xi)(V_A^*V_B)_{r,r+\eta}\big|=1$,  if and only if $\beta_2^{-1}\alpha_2\oplus \beta_2^{-1}\delta_2\alpha_1^2\oplus\alpha_1\gamma_1\neq0$, if and only if
$\beta_2^{-1}\alpha_2+\beta_2^{-1}\delta_2\alpha_1^2+\alpha_1\gamma_1\neq0$ in $F$, if and only if
$$\beta_2\alpha_1^{-1}(\beta_2^{-1}\alpha_2+\beta_2^{-1}\delta_2\alpha_1^2+\alpha_1\gamma_1)
=\alpha_1^{-1}\alpha_2+\delta_2\alpha_1+\beta_2\gamma_1=trace(A^{-1}B)\neq0.\qquad \qquad \square$$

\begin{lem}   \label{propth3}Let
$A=
\left[\begin{smallmatrix}
  \alpha_1 & \beta_1\\
  \gamma_1&\delta_1
\end{smallmatrix}\right]
$ and $
B=
\left[\begin{smallmatrix}
  \alpha_2 & \beta_2\\
  \gamma_2&\delta_2
\end{smallmatrix}\right]
$ in $SL(2,F)$ such that $\beta_1\neq 0,\beta_2\neq 0$. If $trace(A^{-1}B)\neq0$, then $\Phi_A$ and $\Phi_B$ are mutually unbiased.
\end{lem}

\noindent {\bf Proof} Let $m,n\in F$. Then
$$(V_A)_{m,n}=\frac{1}{\sqrt{q}}\lambda({\beta_1^{-1}}(\alpha_1 n^2+2mn+\delta_1 m^2)), \quad
(V_B)_{m,n}=\frac{1}{\sqrt{q}}\lambda({\beta_2^{-1}}(\alpha_2 n^2+2mn+\delta_2 m^2)).$$
It follow that
$$(V_A^*V_B)_{m,n}=\displaystyle\sum_{k\in F}\overline{(V_A)_{k,m}}(V_B)_{k,n}=\frac{1}{q}\displaystyle\sum_{k\in F}
\lambda({\beta_2^{-1}}(\alpha_2 n^2+2kn+\delta_2 k^2)-{\beta_1^{-1}}(\alpha_1 m^2+2km+\delta_1 k^2))$$
Let $\xi,\eta\in F$. Then we have
$$\begin{array}{lll}(V_A^*V_B)_{r,r+\eta}&=&\dfrac{1}{q}\displaystyle\sum_{k\in F}
\lambda({\beta_2^{-1}}(\alpha_2 (r\oplus \eta)^2+2k(r\oplus \eta)+\delta_2 k^2)-{\beta_1^{-1}}(\alpha_1 r^2+2kr+\delta_1 k^2))
\\&=&\dfrac{1}{q}\displaystyle\sum_{k\in F}
\lambda({\beta_2^{-1}}(\alpha_2 (r+\eta)^2+2k(r+ \eta)+\delta_2 k^2)-{\beta_1^{-1}}(\alpha_1 r^2+2kr+\delta_1 k^2))\\
&=&\dfrac{1}{q}\lambda(\alpha_2\beta_2^{-1}\eta^2)\ \lambda((\beta_2^{-1}\alpha_2-\beta_1^{-1}\alpha_1)r^2+2r\alpha_2\beta_2^{-1}\eta)
\displaystyle\sum_{k\in F}\lambda((\beta_2^{-1}\delta_2-\beta_1^{-1}\delta_1)k^2+2k(\beta_2^{-1}r+\beta_1^{-1}r+\beta_2^{-1}\eta))
\end{array}$$
It follows that
$$\begin{array}{ll}&\displaystyle\sum_{r\in F}\lambda(2r\xi)(V_A^*V_B)_{r,r+\eta}
\\ =&\dfrac{1}{q}\displaystyle\sum_{r\in F}\lambda(2r\xi)\lambda(\alpha_2\beta_2^{-1}\eta^2)\ \lambda((\beta_2^{-1}\alpha_2-\beta_1^{-1}\alpha_1)r^2+2r\alpha_2\beta_2^{-1}\eta)
\displaystyle\sum_{k\in F}\lambda((\beta_2^{-1}\delta_2-\beta_1^{-1}\delta_1)k^2+2k(\beta_2^{-1}r+\beta_1^{-1}r+\beta_2^{-1}\eta))\\
=&\dfrac{1}{q}\lambda(\alpha_2\beta_2^{-1}\eta^2)\displaystyle\sum_{r\in F}\ \lambda((\beta_2^{-1}\alpha_2-\beta_1^{-1}\alpha_1)r^2+2r^2(\alpha_2^2\beta_2^{-2}\eta^2+\xi^2))
\displaystyle\sum_{k\in F}\lambda((\beta_2^{-1}\delta_2-\beta_1^{-1}\delta_1)k^2+2k(\beta_2^{-1}r+\beta_1^{-1}r+\beta_2^{-1}\eta))
\end{array}$$
For the second sum, we have equations:
$$\begin{array}{ll}
&\displaystyle\sum_{k\in F}\lambda((\beta_2^{-1}\delta_2-\beta_1^{-1}\delta_1)k^2+2k(\beta_2^{-1}r+\beta_1^{-1}r+\beta_2^{-1}\eta))\\
=&\displaystyle\sum_{k\in F}\lambda((\beta_2^{-1}\delta_2-\beta_1^{-1}\delta_1)k^2+2k^2(\beta_2^{-2}r^2+\beta_1^{-2}r^2+\beta_2^{-2}\eta^2))\\
=&\displaystyle\sum_{k\in F}\lambda(k((\beta_2^{-1}\delta_2-\beta_1^{-1}\delta_1)+2(\beta_2^{-2}r^2+\beta_1^{-2}r^2+\beta_2^{-2}\eta^2)))\\
=&\displaystyle\sum_{k\in F}\lambda(k((\beta_2^{-1}\delta_2+\beta_1^{-1}\delta_1)+2(\beta_2^{-2}r^2+\beta_1^{-2}r^2+\beta_2^{-2}\eta^2+\beta_1^{-1}\delta_1)))\\
=&\displaystyle\sum_{k\in F}\lambda(k((\beta_2^{-1}\delta_2\oplus\beta_1^{-1}\delta_1)+2((\beta_2^{-2}\oplus\beta_1^{-2})r^2\oplus
\underset{d}{\underbrace{\beta_2^{-2}
 \eta^2\oplus\beta_1^{-1}\delta_1\oplus\beta_2^{-1}\delta_2\beta_1^{-1}\delta_1 }})))
\end{array}$$

\noindent {\bf Case 1} \ $\beta_2^{-1}\delta_2\oplus\beta_1^{-1}\delta_1= 0$.
We claim that  $\beta_1^{-2}\oplus\beta_2^{-2}\neq 0.$

{\it  Proof of the claim: }\  Suppose that
$\beta_1^{-2}\oplus\beta_2^{-2}=0.$   Then we get  equations
$\beta_2^{-1}\delta_2+\beta_1^{-1}\delta_1=0$ and $ \beta_1^{-2}+\beta_2^{-2}= 0$ in $F$.
It follows that $\beta_1=\beta_2$ and $\delta_1=\delta_2$. Since  both $A$ and $B$ are in $SL(2,F)$, we have $\alpha_1\delta_1+\beta_1\gamma_1=\alpha_2\delta_2+\beta_2\gamma_2=1$.
Then  $\alpha_1\delta_1+\beta_1\gamma_1+\alpha_2\delta_2+\beta_2\gamma_2=\beta_1\gamma_2+\beta_2\gamma_1
+\alpha_2\delta_1+\alpha_1\delta_2=trace(A^{-1}B)=0$, which is a contradiction.

 Now we return to calculate $\big|\displaystyle\sum_{r\in F}\lambda(2r\xi)(V_A^*V_B)_{r,r+\eta}\big|$:
 $$\begin{array}{lll}
 \big|\displaystyle\sum_{r\in F}\lambda(2r\xi)(V_A^*V_B)_{r,r+\eta}\big|&=&\dfrac{1}{q}\big|\displaystyle\sum_{r\in F}\ \lambda((\beta_2^{-1}\alpha_2-\beta_1^{-1}\alpha_1)r^2+2r^2(\alpha_2^2\beta_2^{-2}\eta^2+\xi^2))
\displaystyle\sum_{k\in F}\lambda(2 k((\beta_2^{-2}\oplus\beta_1^{-2})r^2\oplus d ))\big|\\
&=&|\lambda((\beta_2^{-1}\alpha_2-\beta_1^{-1}\alpha_1)r^2+2r^2(\alpha_2^2\beta_2^{-2}\eta^2+\xi^2))|\qquad (r^2=(\beta_2^{-2}+\beta_1^{-2})^{-1}d\  \mbox{in} \ F)
\\&=&1,
\end{array}$$
 the second equality follows from Lemma \ref{lem21}.

\noindent {\bf Case 2}\ $\beta_2^{-1}\delta_2\oplus\beta_1^{-1}\delta_1\neq  0$.

 Set $u=(\beta_2^{-1}\delta_2\oplus\beta_1^{-1}\delta_1)^{-1}((\beta_2^{-2}\oplus\beta_1^{-2})r^2\oplus d)$.
 It follows from Lemma \ref{lem21} that
 $$\displaystyle\sum_{k\in F}\lambda(k((\beta_2^{-1}\delta_2-\beta_1^{-1}\delta_1)+2(\beta_2^{-2}r^2+\beta_1^{-2}r^2+\beta_2^{-2}\eta^2)))
=\Gamma(1)\lambda(-u).$$
It follows that
$$\begin{array}{lll}\displaystyle\sum_{r\in F}\lambda(2r\xi)(V_A^*V_B)_{r,r+\eta}
%&=&\dfrac{\Gamma(1)}{q}\lambda(\alpha_2\beta_2^{-1}\eta^2)\displaystyle\sum_{r\in F}\ \lambda((\beta_2^{-1}\alpha_2-\beta_1^{-1}\alpha_1)r^2+2r(\alpha_2\beta_2^{-1}\eta+\xi))
%\lambda(-u)\\
&=&\dfrac{\Gamma(1)}{q}\lambda(\alpha_2\beta_2^{-1}\eta^2)\displaystyle\sum_{r\in F}\ \lambda((\beta_2^{-1}\alpha_2-\beta_1^{-1}\alpha_1)r^2+2r^2(\alpha_2^2\beta_2^{-2}\eta^2+\xi^2))
\lambda(-u)\\
&=&\dfrac{\Gamma(1)}{q}\lambda(\alpha_2\beta_2^{-1}\eta^2)\displaystyle\sum_{r\in F}\ \lambda(u+(\beta_2^{-1}\alpha_2+\beta_1^{-1}\alpha_1)r^2+2(u+r^2(\alpha_2^2\beta_2^{-2}\eta^2+\xi^2+\beta_1^{-1}\alpha_1)))
\end{array}$$
Note that $|\Gamma(1)|=\sqrt{q}$.  By the properties of the character $\lambda$, we get that there exists $c\in \mathcal{T}_s$ such that
$$\bigg|\displaystyle\sum_{r\in F}\lambda(2r\xi)(V_A^*V_B)_{r,r+\eta}\bigg|
=\dfrac{1}{\sqrt{q}}\bigg|\displaystyle\sum_{r\in F}\lambda(r((\beta_2^{-1}\delta_2\oplus\beta_1^{-1}\delta_1)^{-1}(\beta_2^{-2}\oplus\beta_1^{-2})\oplus(\beta_2^{-1}\alpha_2\oplus\beta_1^{-1}\alpha_1)+2c))\bigg|.
$$
It follows from Lemmas \ref{lem21}  and \ref{lem22}
 that  $\Phi_A$ and $\Phi_B$ are mutually unbiased if and only if  $\big|\displaystyle\sum_{r\in F}\lambda(2r\xi)(V_A^*V_B)_{r,r+\eta}\big|=1$,  if and only if
 $(\beta_2^{-1}\delta_2\oplus\beta_1^{-1}\delta_1)^{-1}(\beta_2^{-2}\oplus\beta_1^{-2})\oplus(\beta_2^{-1}\alpha_2\oplus\beta_1^{-1}\alpha_1)\neq 0$, if and only if in
$F$ the following equation holds:
$$(\beta_2^{-1}\delta_2+\beta_1^{-1}\delta_1)^{-1}(\beta_2^{-2}+\beta_1^{-2})+(\beta_2^{-1}\alpha_2+\beta_1^{-1}\alpha_1)\neq0.\qquad (*)$$
Note that   $(1+\alpha_1\delta_1=\beta_1\gamma_1)$ and $(1+\alpha_2\delta_2=\beta_2\gamma_2)$.
Then $(*)$ holds if and only if
$$\beta_2^{-2}+\beta_1^{-2}\neq (\beta_2^{-1}\delta_2+\beta_1^{-1}\delta_1)(\beta_2^{-1}\alpha_2+\beta_1^{-1}\alpha_1)
=\beta_2^{-2}\alpha_2\delta_2+\beta_1^{-2}\alpha_1\delta_1+\beta_1^{-1}\beta_2^{-1}(\alpha_2\delta_1+\alpha_1\delta_2)$$
if and only if   \hskip 2cm
$\beta_2^{-2}(1+\alpha_2\delta_2)+\beta_1^{-2}(1+\alpha_1\delta_1)
\neq\beta_1^{-1}\beta_2^{-1}(\alpha_2\delta_1+\alpha_1\delta_2),$
\\
if and only if \hskip 2cm
$\beta_2^{-1}\gamma_2+\beta_1^{-1}\gamma_1
\neq\beta_1^{-1}\beta_2^{-1}(\alpha_2\delta_1+\alpha_1\delta_2),$
\\
if and only if \hskip 2cm
$\beta_1\gamma_2+\beta_2\gamma_1
\neq\alpha_2\delta_1+\alpha_1\delta_2,$
\\
if and only if \hskip 2cm
$trace(A^{-1}B)=\beta_1\gamma_2+\beta_2\gamma_1
+\alpha_2\delta_1+\alpha_1\delta_2\neq0$. \qquad $\square$

\medskip

By Lemmas \ref{propth1},\ref{propth2} and \ref{propth3}, we get the following proposition.
\begin{prop} \label{propall}Let
$A=
\left[\begin{smallmatrix}
  \alpha_1 & \beta_1\\
  \gamma_1&\delta_1
\end{smallmatrix}\right]
$ and $
B=
\left[\begin{smallmatrix}
  \alpha_2 & \beta_2\\
  \gamma_2&\delta_2
\end{smallmatrix}\right]
$ in $ SL(2,F)$. If $trace(A^{-1}B)\neq0$, then $\Phi_A$ and $\Phi_B$ are mutually unbiased.
\end{prop}
\begin{Def}\label{def1} A non-empty subset $\mathcal{C}$ of $ SL(2,F)$ is called
a {\bf trace-zero excluded subset} if $$\forall \   A,B\in \mathcal{C}, \quad  trace(A^{-1}B)= 0\Longleftrightarrow A=B.$$
\end{Def}

As a direct consequence of Proposition \ref{propall} and Definition \ref{def1}, we get the following theorem.

\begin{theo}\label{theo2}
Suppose that $\mathcal{C}$ is a trace-zero excluded subset of $SL(2,F)$.
Then $(\Phi_{A})_{A\in \mathcal{C}}$ is a  set of MUMEBs in $\mathbb{C}^q\otimes\mathbb{C}^q$.
\end{theo}

\begin{rem} Given two unitary matrices $A$ and $B$ in $M_q(\mathbb{C})$, one can check whether $\Phi_A$ and $\Phi_B$ are mutually unbiased by Lemma \ref{lem22}. However, finding suitable unitary matrices  becomes more and more difficult  with
the increase of the dimension $q$. The theorem provides a new idea to construct MUMEBs in bipartite system, because it not only simplifies  the calculation but also provides a large number of  unitary matrices.
\end{rem}
\subsection{The constructions and examples}

Let $A=\left[\begin{smallmatrix}
  a_1& c_1 \\
    c_1&  b_1
 \end{smallmatrix}\right]$ and
 $B=\left[\begin{smallmatrix}
  a_2& c_2 \\
    c_2&  b_2
 \end{smallmatrix}\right]$ in $SL(2,F)$ such that $a_1\neq b_1,a_2\neq b_2$.
  Then trace$(A^{-1}B)=a_2b_1+a_1b_2$. Thus, trace$(A^{-1}B)\neq 0$ if and only if $(a_1,b_1)$ and $(a_2,b_2)$
  are linearly independent. It follows that we can always construct trace-zero excluded subsets of cardinality $q+1$. In particular, we get the following result.

\begin{prop}\label{propex1} Let $\xi$ be a primitive element in $F$. Then the set
$$\mathcal{C}=\bigg\{
\left[\begin{smallmatrix}
  1& 1 \\
   1&  0
 \end{smallmatrix}\right],
 \left[\begin{smallmatrix}
  0& 1\\
   1&  1
 \end{smallmatrix}\right]\bigg\}\
 \cup\  \bigg\{\left[\begin{smallmatrix}
  1& \sqrt{1+\xi^k} \\
   \sqrt{1+\xi^k}&  \xi^k
 \end{smallmatrix}\right]\mid 0\leq k\leq q-2\bigg\}$$ is a trace-zero excluded subset.
 %In particular, we get a new set of MUMEBs consisting of $q+1$ elements.
\end{prop}

The following result provides  a completely new set of MUMEBs in  $\mathbb{C}^{2^s}\otimes \mathbb{C}^{2^s}$.
\begin{prop}\label{propex2} Let $\xi$ be a primitive element in ${F}$.
 Then the set
$$\mathcal{C}= \bigg\{A_k=\left[\begin{smallmatrix}
 \xi^k& 0 \\
    0 &  \xi^{-k}
 \end{smallmatrix}\right], \quad
 B_k=\left[\begin{smallmatrix}
 \xi^k& \xi^k \\
   \xi^{-k}  &0
 \end{smallmatrix}\right], \quad
 C_k=\left[\begin{smallmatrix}
 0& \xi^k \\
   \xi^{-k} &  \xi^{-k}
 \end{smallmatrix}\right]
\quad \mid \quad 0\leq k\leq q-2\bigg\}$$
is a trace-zero excluded subset.  In particular, $M(q,q)\geq 3(q-1).$
 \end{prop}

{\bf Proof}  Let $0\leq k,j\leq q-2$.  Then trace$(A_k^{-1}B_j)=$trace$(B_k^{-1}C_j)=\xi^{j-k}\neq 0$ and trace$(A_k^{-1}C_j)=\xi^{k-j}\neq 0.$  Moreover, if $k\neq j$, then trace$(B_k^{-1}B_j)=$trace$(C_k^{-1}C_j)=\xi^{k-j}+\xi^{j-k}\neq 0.$
Consequently, $\mathcal{C}$ is a trace-zero excluded subset.  It follows from Theorem
\ref{theo2} that $M(q,q)\geq 3(q-1).$
\qquad  $\square$

\begin{rem} Proposition \ref{propex2} is a generalization of  \cite[Theorem 4.8]{xu}, where it was proved that $M(q,q)\geq 2(q-1)$
\end{rem}

%Similarly, we get the following.
%
%
%\begin{prop}\label{prop67} Let $\xi$ be a primitive element in ${F}$.
% Then the set
%$$\mathcal{C}= \bigg\{\left[\begin{smallmatrix}
% \xi^k& 0 \\
%    0 &  \xi^{-k}
% \end{smallmatrix}\right], \quad
%\left[\begin{smallmatrix}
% \xi^k& \xi^{-k} \\
%   \xi^{k}  &0
% \end{smallmatrix}\right], \quad
% \left[\begin{smallmatrix}
% 0& \xi^{-k} \\
%   \xi^{k} &  \xi^{-k}
% \end{smallmatrix}\right]
%\quad \mid \quad 0\leq k\leq q-2\bigg\}$$
%is a trace-zero excluded subset.
% \end{prop}
%
%

In the last of this section, we give   examples to show that our construction is different from the wok in \cite{xu2} and  how the main results can be used to construct MUMEBs.

Let $h_2(x)=x^2+x+1$ over $\mathbb{Z}_2[X]$. It is easy to check that the unique polynomial $h(x)\in \mathbb{Z}_4[X]$ satisfying $h(x)\equiv h_2(x)$ (mod $2$) and $h(x)$ divides $x^{2^s-1}-1$ (mod $ 4$)  is $h(x)=x^2+x+1$.
Let $\xi$ be a root of $h(x)$ such that $\xi^{3}=1$ and $\alpha=\phi(\xi)$.
Then we have the following:   $$\mathbb{F}_4=\{0,1,\alpha,\alpha^2=1+\alpha\},\quad
\mathcal{T}_2=\{0,1,\xi,\xi^2\}, \quad  \mathbb{Z}_4[\xi]=\{a+2b\mid a,b\in \mathcal{T}_2\}.$$
By a direct calculation, we get the following:
$$\begin{array}{llll}
tr(0)=0& tr(1)=2 & tr(\xi)=3 & tr(\xi^2)=3\\
tr(2)=0& tr(1+2)=2&tr(\xi+2)=3& tr(\xi^2+2)=3\\
tr(2\xi)=2&tr(1+2\xi)=0&tr(\xi+2\xi)=1&tr(\xi^2+2\xi)=1\\
tr(2\xi^2)=2&tr(1+2\xi^2)=0&tr(\xi+2\xi^2)=1&tr(\xi^2+2\xi^2)=1
\end{array}$$

\begin{rem} Set $A=\left[\begin{smallmatrix}
1&1\\1&0\end{smallmatrix}\right]$.
We get that  $A^2=\left[\begin{smallmatrix}
0&1\\1&1\end{smallmatrix}\right]$ and
$$V_A=\dfrac{1}{2}\left[\begin{smallmatrix}
                     1 & -1 & -i & -i\\
                     1 &-1 & i & i \\
                     1 & 1& i& -i \\
                    1 & 1 & -i & i\end{smallmatrix}\right]
                   \qquad
                  V_A{^2}=\dfrac{1}{2}\left[\begin{smallmatrix}
                     -i & -i & -i & -i\\
                     i &i & -i & -i \\
                    1 & -1 & -1& 1\\
                    1&-1&1&-1
                     \end{smallmatrix}\right]\qquad
                    V_{A^2}=\left[\begin{smallmatrix}
                     1 & 1 & 1 & 1\\
                     -1 &-1 & 1 & 1 \\
                     -i & i& i& -i \\
                    -i & i & -i & i
                   \end{smallmatrix}\right]$$
It follows that there does not exist $z\in \mathbb{C}$  with $|z|=1$ such that  $V_{A^2}=zV_A^2$, which is different from
\cite[Lemma 3.2]{xu2}, where it was proved that for each $A\in SL(2,F)$ with char$(F)\neq 2$, there exists $z\in \mathbb{C}$  with $|z|=1$ such that  $V_{A^2}=zV_A^2$.
\end{rem}

%%%%%%%%%%%%%%%%%%%%%%%%%%

\begin{exe} Set $s=2$ in Proposition \ref{propex1}. Set
$$
 D_k=\left[\begin{smallmatrix}
  1& \sqrt{1+\xi^k} \\
   \sqrt{1+\xi^k}&  \xi^k
 \end{smallmatrix}\right]\ (0\leq k\leq 2) \qquad
 D_3=\left[\begin{smallmatrix}
  1& 1 \\
   1&  0
 \end{smallmatrix}\right] \qquad
D_4= \left[\begin{smallmatrix}
  0& 1\\
   1&  1
 \end{smallmatrix}\right] $$
 By a direct calculation, we  get the following unitary matrices:
$$\begin{array}{lllll}V_{D_0}=I_4 &
V_{D_1}=\dfrac{1}{2}\left[\begin{smallmatrix}
                     1 & -i & -i & -1\\
                     -1 &-i & i & -1 \\
                     -i & -1& 1& -i \\
                    -i & 1 & 1 & i\end{smallmatrix}\right]
                    &
                    V_{D_2}= \dfrac{1}{2}\left[\begin{smallmatrix}
                     1 & -i & -1 & -i\\
                     -1 &-i & -1 &i \\
                     -i &1& i& 1 \\
                    -i&-1 & -i & 1
                   \end{smallmatrix}\right]
                   &
                     V_{D_3}=\dfrac{1}{2}\left[\begin{smallmatrix}
                     1 & -1 & -i & -i\\
                     1 &-1 & i & i \\
                     1 & 1& i& -i \\
                    1 & 1 & -i & i\end{smallmatrix}\right]
                   &
                   V_{D_4}= \dfrac{1}{2}\left[\begin{smallmatrix}
                     1 & 1 & 1 & 1\\
                     -1 &-1 & 1 & 1 \\
                     -i & i& i& -i \\
                    -i & i & -i & i
                   \end{smallmatrix}\right]

\end{array}$$
By Theorem \ref{theo2},  $(\Phi_{V_{D_i}})_{0\leq i\leq 4}$ is a set of MUMEBs  in $\mathbb{C}^4\otimes \mathbb{C}^4$.
\end{exe}

\begin{exe} Set $s=2$ in Proposition \ref{propex2}. By a direct calculation, we  get the following unitary matrices:
$$\begin{array}{lll}V_1=V_{A_0}=\quad I_4
&\quad
 V_2= V_{A_1}=\quad \left[\begin{smallmatrix}
                     1 & 0 & 0 & 0\\
                     0 &0 & 0 & 1 \\
                     0& 1& 0& 0 \\
                    0& 0& 1 & 0
                   \end{smallmatrix}\right]
&\quad
 V_3= V_{A_2}=\quad \left[\begin{smallmatrix}
                     1 & 0 & 0 & 0\\
                     0 &0 & 1 & 0\\
                     0& 0& 0& 1 \\
                    0& 1& 0 & 0
                   \end{smallmatrix}\right]
\\ \\
V_4= V_{B_0}=\dfrac{1}{2}\left[\begin{smallmatrix}
                     1 & -1 & -i & -i\\
                     1 &-1 & i & i \\
                     1 & 1& i& -i \\
                    1 & 1 & -i & i\end{smallmatrix}\right]
                    &\quad
                   V_5= V_{B_1}= \dfrac{1}{2}\left[\begin{smallmatrix}
                     1 & -1 & -i & -i\\
                     1 &1 & -i &i \\
                     1 &-1& i& i \\
                    1&1 & i & -i
                   \end{smallmatrix}\right]
                   &\quad
                   V_6=  V_{B_2}= \dfrac{1}{2}\left[\begin{smallmatrix}
                     1 & -1 & -i & -i\\
                     1 &1 & i & -i \\
                     1 & 1&-i&i \\
                    1 & -1 & i & i
                   \end{smallmatrix}\right]
                   \\ \\
                 V_7= V_{C_0}= \dfrac{1}{2}\left[\begin{smallmatrix}
                     1 & 1 & 1 & 1\\
                     -1 &-1 & 1 & 1 \\
                     -i & i& i& -i \\
                    -i & i & -i & i
                   \end{smallmatrix}\right]
                   &\quad
V_8=V_{C_1}= \dfrac{1}{2}\left[\begin{smallmatrix}
                     1 & 1 & 1 & 1\\
                     -i &i & -i & i \\
                     -1& -1& 1& 1 \\
                    -i & i &i & -i
                   \end{smallmatrix}\right]
                   &\quad
V_9= V_{C_2}= \dfrac{1}{2}\left[\begin{smallmatrix}
                     1 & 1 & 1 & 1\\
                     -i &i &i & -i \\
                     -i&i& -i& i \\
                    -1 & -1 &1 &1
                   \end{smallmatrix}\right]
\end{array}$$
By Theorem \ref{theo2},  $(\Phi_{V_i})_{1\leq i\leq 9}$ is a set of MUMEBs  in $\mathbb{C}^4\otimes \mathbb{C}^4$.
\end{exe}

\section{Conclusions}
 To construct mutually unbiased maximally entangled bases in bipartite  system $\mathbb{C}^{2^s}\otimes \mathbb{C}^{2^s}$, we introduce the notation of trace-zero excluded subset of  $SL(2,\mathbb{F}_{2^s})$ and establish a  relation between trace-zero excluded subsets of $SL(2,\mathbb{F}_{2^s})$ and MUMEBs  in $\mathbb{C}^{2^s}\otimes \mathbb{C}^{2^s}$. We obtain  a set of MUMEBs in $\mathbb{C}^{2^s}\otimes \mathbb{C}^{2^s}$ with  cardinality $3(q-1)$
 by constructing  trace-zero excluded subsets in $SL(2,\mathbb{F}_{2^s})$, which generalizes  one of the main  results   in \cite{xu}.

In the paper, we  provide a new method to construct MUMEBs in $\mathbb{C}^{2^s}\otimes\mathbb{C}^{2^s}$. However, the  trace-zero excluded subsets  constructed in the paper are limited.  It would be interesting to construct  trace-zero excluded subsets of  $SL(2,\mathbb{F}_{2^s})$ with larger cardinalities, which will raise the lower bound of $M(2^s,2^s)$.
\medskip

%\noindent {\bf Acknowledgement}\quad  The author thanks
% the anonymous referees  for helpful comments.

\bigskip 

2019.10
\end{document}